\documentclass[aps,twocolumn,pra,showpacs]{revtex4-1}
\usepackage{amssymb}
\usepackage{mathrsfs}
\usepackage{graphicx}
\usepackage{amsmath,amssymb,amsthm,bbm,latexsym}
\usepackage{amsthm}
\usepackage{amsbsy}
\usepackage{epsfig}
\usepackage{graphicx}
\usepackage{float}
\usepackage{dcolumn}
\usepackage{amsmath}

\begin{document}

\author{Zhao-Ming Wang$^{1}$\thanks{%
Email address: mingmoon78@126.com}, Rui-Song Ma$^{1,2}$, C. Allen Bishop$^{3}$%
, and Yong-Jian Gu$^{1}$\thanks{%
Email address: yjgu@ouc.edu.cn} }
\affiliation{$^{1}$ Department of Physics, Ocean University of China, Qingdao, 266100,
China}
\affiliation{$^{2}$ Institute of Physics, Chinese
Academy of Sciences, Beijing, 100080, China}
\affiliation{$^{3}$ Department of Physics, Southern Illinois University, Carbondale,
Illinois 62901-4401, USA}
\title{Quantum state transfer through a spin chain in a multi-excitation
subspace}

\begin{abstract}
We investigate the quality of quantum state transfer through a uniformly
coupled antiferromagnetic spin chain in a multi-excitation subspace. The
fidelity of state transfer using multi-excitation channels is found to
compare well with communication protocols based on the ground state of a
spin chain with ferromagnetic interactions. Our numerical results support
the conjecture that the fidelity of state transfer through a
multi-excitation subspace only depends on the number of initial excitations
present in the chain and is independent of the excitation ordering. Based on
these results, we describe a communication scheme which requires little
effort for preparation.
\end{abstract}

\maketitle

\section{Introduction}

Spin qubits have been considered in many quantum communication analyses due
to their wide applicability in various solid-state devices. A prominent form
of interaction between the spins is the exchange coupling which can be
described by the Heisenberg model. The Heisenberg spin chain has been
extensively studied as a communication channel for many quantum information
processing tasks \cite{BOSE2}. In most theoretical treatments of the subject
one assumes a ferromagnetic (FM) coupling and the channel spins are assumed
to be initialized to the completely polarized ground state. However, most
physical realizations of quantum spin channels have an antiferromagnetic
(AFM) ordering \cite{Hirjibehedin2006}. The ground state wave function for
the antiferromagnetic XY Hamiltonian contains numerous amplitudes within a
multi-excitation subspace. Although the ground state configuration is more
complicated than the FM ground state, it may prove to be a more suitable
pathway for quantum communication. It appears that the first proposal for
using antiferromagnetic spin chains for quantum state transfer was provided
in \cite{Camposprl2007}. There it was shown that AFM Heisenberg chains can
represent good quantum channels for robust finite temperature teleportation
and state transfer.

Recently, an experimental proposal for the quantum simulation of an AFM spin
chain in an optical lattice has been provided \cite{Simon2011}. It has also
been shown that the ground state of some one-dimensional spin models with
finite correlation length can distribute entanglement between long distance
sites \cite{Campos2006}. As the chain length increases, true long-distance
entanglement, characterized by energy gaps above the ground state, vanishes
exponentially. However, long distance entanglement can be supported by the
ground state of spin models with infinite correlation length defined on
one-dimensional open chains with small end bonds \cite{Campos2006}. Open
quantum spin chains endowed with XY-like Hamiltonians containing
nearest-neighbor interactions have also investigated \cite{Campospra2007}.
For dimerized XY chains, true long distance entanglement has been found to
exist only at zero temperature although "quasi long-distance" entanglement
can be realized in open XY chains with small end bonds \cite{Campospra2007}.
There it was found that the entanglement properties slowly fall off with the
size of the chain and that efficient qubit teleportation can be realized
with high fidelity in long chains even at moderately low temperatures.

Quantum computing is possible using a wide variety of systems assembled from
antiferromagnetically coupled spins \cite{Meier2003}. Entanglement
properties in two-dimensional AFM models have been studied \cite%
{Roscilde2005} and it has been shown that multiqubit entanglement can be
generated efficiently via a quantum data bus consisting of spin chains with
strong static AFM couplings \cite{Friesen2007,Oh20101}. The effects of
fluctuating exchange couplings and magnetic fields on the fidelity of data
bus transfer have also been investigated \cite{Oh20102}. Bayat \textit{et al.%
} studied the entanglement transfer through an AFM spin chain and found that
when compared to the FM case, the entanglement can be transmitted faster,
with less decay, and with a much higher purity \cite{Bayat2010}.
Furthermore, Wang \textit{et al}. demonstrated that near-perfect
entanglement can be generated between the first and last spins of an AFM
isotropic Heisenberg chain by applying a magnetic field to a single site in
a specific direction \cite{Wang2010}. Moreover, perfect state transfer
across a strongly coupled AFM spin chain or ring has been shown to be
possible using weakly coupled external qubits \cite{Oh2011}. Detrimental
dispersion effects on the transmission are found to be strongly reduced by
modifying only one or two bonds in an XX spin chain and a transmission
fidelity more than $99\%$ for arbitrary long chains is gained \cite%
{Apollaro2012,Banchi2011}.

It is known that quantum information propagates dispersively through most
spin chains due to the nontrivial structure of the many-body Hamiltonian
that describes the channel \cite{Bayat2011}. Designing a non-dispersive
channel requires the intricate engineering of the local couplings \cite%
{Christandl2004}. Dispersion is always detrimental to the information
transmission, usually there is always some portion of information left in
the chain after measurement and hence lost to the receiver \cite{Wiesniak}.
Although perfect state transfer cannot typically be achieved using uniformly
coupled chains alone, the investigation of state transfer through chains of
this sort is warranted due to the relative ease of preparation compared to
more elaborate schemes.

In this work, we compare the quality of state transmission for several
initial configurations of a spin chain. Specifically, we examine spin chains
initialized to the N\'{e}el state (AFM arrangements) and find that the
average fidelity of state transfer using these channels is similar to that
which can be obtained using the completely polarized FM ground state. The
results follow the original proposal of Bose \cite{Bose2003}. Higher average
fidelities occur using this AFM arrangement for chains having an appropriate
length, and in some cases these higher values are accompanied by shorter
arrival times as well. It is also found that the quality of transmission
through chains prepared in a multi-excitation state which contain a fixed
number of initial excitations remains nearly identical regardless of the
order in which these excitations occur. These results bode well for
potential realizations of these communication channels as they suggest a
simplification of the initialization process. We analyze the magnetic field
dependence of the fidelity in all cases and find that this measure is
influenced strongly by fluctuations in the external field.

\begin{figure}[tbh]
\centering
\includegraphics[scale=0.8,angle=0]{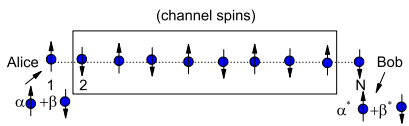}
\caption{(Color online) Schematic of our spin chain communication channel.
After initializing the chain to the N\'{e}el state Alice encodes an
arbitrary qubit state at one end and allows it to propagate freely to the
other end.}
\label{fig:1}
\end{figure}

We first consider the N\'{e}el state configuration shown schematically in
Fig. 1. We will provide our mathematical model next and derive expressions
for the fidelity measure. In Sec. III we discuss our results for this and
other AFM arrangements and compare these results to the FM case. Sec. IV
concludes with a summary of our findings.

\section{The model}

We consider a linear spin chain with uniform nearest-neighbor XY couplings.
The Hamiltonian is given by

\begin{equation}
H=\frac{J}{2}[\sum_{l=1}^{N-1}(\sigma _{l}^{x}\sigma _{l+1}^{x}+\sigma
_{l}^{y}\sigma _{l+1}^{y})-\frac{2h}{J}\sum_{l=1}^{N}\sigma _{l}^{z}]
\end{equation}%
In this expression \textit{J} denotes the exchange constant between adjacent
spins. For FM chain, $J<0$ and for AFM chain, $J>0$. \textit{h} represents
the external magnetic field strength of a field applied along the \textit{z}
direction, and $\sigma _{l}^{x,y,z}$ signifies the Pauli operators acting on
spin \textit{l}. We take $\hbar =1$ throughout. Note that for this
Hamiltonian the $z$-component of the total spin $\sigma ^{z}=\sum \sigma
_{l}^{z}$ is a conserved quantity, i.e. $[H,\sigma ^{z}]=0$, which indicates
that the system contains a fixed number of excitations. 
The number of excitations in the chain corresponds to the number of \ $%
\left\vert 1\right\rangle $'s appearing in the state vector, where $%
\left\vert 0\right\rangle $ represents the spin-down state of a spin qubit
and $\left\vert 1\right\rangle $ represents a spin-up state. For instance,
the state $\left\vert 111...1\right\rangle $ labels a chain containing
\textit{N} excitations.

This Hamiltonian can be diagonalized by means of the Jordan-Wigner
transformation which maps spins to one dimensional spinless fermions with
creation operators defined by $c_{k}^{\dagger
}=(\prod\limits_{l=1}^{k-1}-\sigma _{l}^{z})\sigma _{k}^{+}$. Here $\sigma
_{k}^{\pm}=\frac{1}{2}(\sigma _{k}^{x}\pm i\sigma _{k}^{y})$ denotes the
spin raising and lowering operations at site \textit{k.} The action of $%
c_{k}^{\dagger }$ is to flip the spin at site \textit{k} from down to up.
For indices \textit{l} and \textit{m} the operators $c_{l}$ and $%
c_{m}^{\dagger }$ satisfy the anticommutation relation $\{c_{l},c_{m}^{%
\dagger }\}=\delta _{lm}$. The $z$ component of the total spin is a
conserved quantity and thus the total number of excitations $M = \sum_{k}
c_{k}^{\dagger }c_{k}$ in the chain remains constant.

The time dependence of the operator $c_{k}^{\dagger }$ has been calculated
\cite{Amico} and is given by

\begin{equation}
c_{k}^{\dagger }(t)=\sum_{l=1}^{N}f_{k,l}(t)c_{l}^{\dagger }.
\end{equation}%
In this expression the transition amplitudes $f_{k,l}$ evolve according to
the relation
\begin{equation}
f_{k,l}(t)=\frac{2}{N+1}\sum_{m=1}^{N}\sin (q_{m}k)\sin (q_{m}l)e^{-iE_{m}t},
\end{equation}%
where $q_{m}=\pi m/(N+1)$, $E_{m}=2h+2J\cos q_{m}.$ In what follows, let us
define the completely polarized state $\left\vert 000\hdots0\right\rangle $
to be $\left\vert \mathbf{0}\right\rangle \equiv \left\vert 000\hdots%
0\right\rangle $ and let $S$ denote a set of $M$ different numbers from $%
1,2,...,N$. The set $S=\{k_{1},k_{2},...,k_{M}\}$ serves to label the sites
where the $M$ excitations initially exist. In this notation, we can express
our initial chain configuration as
\begin{equation}
\left\vert \Psi (0)\right\rangle =\left( \prod\limits_{k\in S}c_{k}^{\dagger
}\right) \left\vert \mathbf{0}\right\rangle .
\end{equation}%
In this work we will only consider spin channels which are initialized to
single ket states of this form. This initial state then evolves to \cite%
{Wichterich}

%
%
%
%

\begin{equation}
\left\vert \Psi(t)\right\rangle =\sum_{l_{1}<l_{2}<...<l_{M}}\det
(A)\left(\prod\limits_{m=1}^{M}c_{l_{m}}^{\dagger }\right)\left\vert\mathbf{0%
} \right\rangle ,
\end{equation}
where
\begin{equation}
A=\left\vert
\begin{array}{cccc}
f_{k_{1},l_{1}} & f_{k_{1},l_{2}} & ... & f_{k_{1},l_{M}} \\
f_{k_{2},l_{1}} & f_{k_{2},l_{2}} & ... & f_{k_{2},l_{M}} \\
... & ... & ... & ... \\
f_{k_{M},l_{1}} & f_{k_{M},l_{2}} & ... & f_{k_{M},l_{M}}%
\end{array}%
\right\vert .
\end{equation}
Note that we have suppressed the explicit time dependence of the amplitudes $%
f_{i,j}(t)$ in the determinant above. The indices $l_1,l_2,...l_M$ have a
similar meaning to the $k_i$ and mark the sites where the excitations have
spread to. We are interested in the fidelity of the state which Bob receives
at site $N$ so we will also need an expression for the reduced density
matrix at this site. This operator has the form \cite{Osborne}
\begin{equation}
\rho _{N}(t)=\left(
\begin{array}{cc}
\left\langle \sigma _{N}^{+}\sigma _{N}^{-}\right\rangle & \left\langle
\sigma _{N}^{-}\right\rangle \\
\left\langle \sigma _{N}^{+}\right\rangle & \left\langle \sigma
_{N}^{-}\sigma _{N}^{+}\right\rangle%
\end{array}%
\right)
\end{equation}
where the symbol $\left\langle {\cdot}\right\rangle $ represents the average
value of $\left\langle \Psi _{l}^{M}(t)\right\vert \cdot\left\vert \Psi
_{l}^{M}(t)\right\rangle $.

Let us assume that the interaction between spins 1 and 2 can be turned on or
off. First the interaction is turned off and Alice prepares an arbitrary
qubit state $\alpha \left\vert 0\right\rangle +\beta \left\vert
1\right\rangle $ at the first site of the chain. Now suppose the channel
which includes spins 2 to \emph{N} is prepared in the N\'{e}el state. The
initial state of the system is given by

\begin{equation}
\left\vert \Psi (0)\right\rangle = \left\{%
\begin{array}{ll}
(\alpha \left\vert 0\right\rangle +\beta \left\vert 1\right\rangle )\otimes
\left\vert 1010...10\right\rangle, & \mbox{for odd }N \\
(\alpha \left\vert 0\right\rangle +\beta \left\vert 1\right\rangle )\otimes
\left\vert 1010...01\right\rangle, & \mbox{for even }N%
\end{array}%
\right.
\end{equation}

Now turn on the interaction and let the system evolve freely. At an
appropriate time Bob will receive a state which resembles the one Alice
prepared. An ideal transmission would result when Bob receives a state which
is identical to the one Alice prepared, in that case the fidelity would be
perfect. Then%
\begin{eqnarray}
\left\vert \Psi (t)\right\rangle &=&\alpha \sum_{l_{1}<l_{2}<...<l_{M_{1}}}
A_{1}\left(\prod\limits_{m=1}^{M_{1}}c_{l_{m}}^{\dagger }\right)\left\vert
\mathbf{0} \right\rangle \   \notag \\
&+&\;\beta \sum_{l_{1}<l_{2}<...<l_{M_{2}}}
A_{2}\left(\prod\limits_{m=1}^{M_{2}}c_{l_{m}}^{\dagger }\right)\left\vert
\mathbf{0}\right\rangle.  \notag \\
\end{eqnarray}
In the expression above we have
\begin{equation}
M_{1}=\frac{N-1}{2}, \;M_{2}=\frac{N+1}{2}, \;\;{\text{for odd }}N
\end{equation}
and
\begin{equation}
M_{1}=\frac{N}{2}, \;M_{2}=\frac{N}{2}+1, \;\;\;\;\;\;\;\;\;{\text{for even }%
}N
\end{equation}
The determinants $A_{1}$ and $A_{2}$ have the same form as $A$ except that $%
M $ is replaced with $M_{1}$ and $M_{2}$ respectively. Now the matrix
elements in Eq. (7) can be calculated as

\begin{eqnarray}
\left\langle \sigma _{N}^{+}\sigma _{N}^{-}\right\rangle &=&\left\vert
\alpha \right\vert ^{2}\Gamma _{1}+\left\vert \beta \right\vert ^{2}\Gamma
_{2},  \notag \\
\left\langle \sigma _{N}^{-}\sigma _{N}^{+}\right\rangle &=&\left\vert
\alpha \right\vert ^{2}\Gamma _{3}+\left\vert \beta \right\vert ^{2}\Gamma
_{4},  \notag
\end{eqnarray}
and
\begin{equation}
\left\langle \sigma _{N}^{+}\right\rangle =\alpha \beta ^{\ast }\Gamma
_{5},\;\;\; \left\langle \sigma _{N}^{-}\right\rangle =\left\langle \sigma
_{N}^{+}\right\rangle ^{\ast }
\end{equation}
with
\begin{eqnarray}
\Gamma _{1} &=& \;\;\; \sum_{l_{1}<l_{2}<...<(l_{M_{1}}=N)}\;\;\;\;(\det
A_{1})^{\ast }\det A_{1},  \notag \\
\Gamma _{2} &=& \;\;\; \sum_{l_{1}<l_{2}<...<(l_{M_{2}}=N)}\;\;\;\;(\det
A_{2})^{\ast }\det A_{2},  \notag \\
\Gamma _{3} &=& \;\;\; \sum_{l_{1}<l_{2}<...<(l_{M_{1}}\neq N)}\;\;\;\;(\det
A_{1})^{\ast }\det A_{1},  \notag \\
\Gamma _{4} &=& \;\;\; \sum_{l_{1}<l_{2}<...<(l_{M_{2}}\neq N)} \;\;\;\;
(\det A_{2})^{\ast }\det A_{2},  \notag \\
\Gamma _{5} &=&\sum_{l_{1}<l_{2}<...<l_{M_{1}}<(l_{M_{2}}=N)}(\det
A_{2})^{\ast }\det A_{1}.
\end{eqnarray}

Let $\alpha =\cos \frac{\theta }{2}$ and $\beta =\sin \frac{\theta }{2}%
e^{i\varphi }$, the average fidelity of transmission at Bob's end can then
be calculated by integration over the unit sphere

\begin{eqnarray}
F &=&\frac{1}{4\pi }\int \left\langle \varphi _{in}\right\vert \rho
_{N}(t)\left\vert \varphi _{in}\right\rangle d\Omega  \notag \\
&=&\frac{1}{3}[\Gamma _{2}+\Gamma _{3}+{\text{Re}}(\Gamma _{5})]+\frac{1}{6}%
[\Gamma _{1}+\Gamma _{4}].
\end{eqnarray}

When two excitations exist (\emph{M}=2) the fidelity can be found to agree
with Eq. (7) of Ref \cite{Wang2011} where we discuss duplex quantum
communication in a FM spin chain.

Here we define the perfect state transfer as "Bob receives a state which is
identical to the one Alice prepared". However, other works\cite%
{BOSE2,Bose2003} about the definition of perfect state transfer allow the
receiver's state has an arbitrary phase factor compared with the sender's
state. In the latter definition, the third term in Eq.(14) is changed to $%
(\left\vert \Gamma _{5}\right\vert \cos \gamma)/3 $, where $\gamma
=\arg \{\Gamma _{5}\}$. To maximize the average fidelity, the magnetic field
must be properly chosen such that $\cos \gamma =1$.

\begin{figure}[tbph]
\centering
\includegraphics[scale=0.7,angle=0]{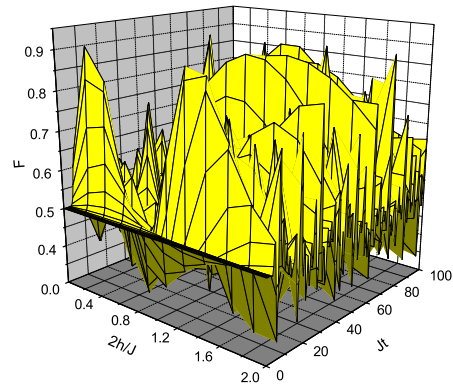}
\caption{(Color online) Plot of the average fidelity of state transfer as a
function of scaling time $Jt$ and scaling external field $2h/J$. The figure
corresponds to a \emph{N}=10 site chain which has been initialized to the N%
\'{e}el state.}
\label{fig:2}
\end{figure}

When $M=1$, i.e. only one excitation exists in the chain, the excited
state contributes as a term in Alice's encoded qubit state $(k_{1}=1)$. In
this case we have $\Gamma _{1}=0$, $\Gamma _{2}=\left\vert
f_{1,N}\right\vert ^{2}$, $\Gamma _{3}=1$, $\Gamma _{4}=1-\left\vert
f_{1,N}\right\vert ^{2}$, and $\Gamma _{5}=f_{1,N}^{\ast }$. The
corresponding average fidelity can be simplified to
\begin{equation}
F=\frac{1}{2}+\frac{1}{3}\left\vert f_{1,N}\right\vert \cos \gamma +\frac{1}{%
6}\left\vert f_{1,N}\right\vert ^{2}.
\end{equation}

Where $\gamma =\arg \{f_{1,N}\}$. This result is in accordance with Bose's
expression (Eq. (6) in Ref \cite{Bose2003}) where the channel spins are
prepared in the ground state of a FM chain.

\section{RESULTS and DISCUSSIONS}

Starting from the initial state (Eq.(8)) the system undergoes a time
evolution described by Eq's. (9)-(12). This evolution can be viewed as the
propagation of the half site excitations. After initialization these
excitations begin to spread outward and at a later time there is typically a
nonzero probability of finding any one of the spins in an excited state.
From the point view of wave mechanics, the transition amplitude $f_{k,l}(t)$
can be viewed as the propagator and the state transfer can be characterized
in terms of the dispersion of all propagators. At some time Tmax maximum
constructive interference occurs at the receiving end and the state of the
spin at Bob's site now has its strongest resemblance to the state which
Alice prepared. First we investigate the effect of a magnetic field on the
evolution of the average fidelity. As an example, Fig.2 illustrates the
average fidelity of transmission through an \emph{N}=10 site chain as a
function of $Jt$ and $2h/J$. We find that the average fidelity changes
abruptly with $Jt$ and $2h/J$ as it oscillates around the value 0.5. The
magnetic field has a pronounced effect on the average fidelity, with
increasing $2h/J$ the oscillation of the average fidelity becomes more
rapid. The maximum average fidelity (MAF) Fmax=0.909 is achieved at $J$Tmax
= 6.0 with $2h/J$=0.2.

\begin{figure}[tbph]
\centering
\includegraphics[scale=1.0,angle=0]{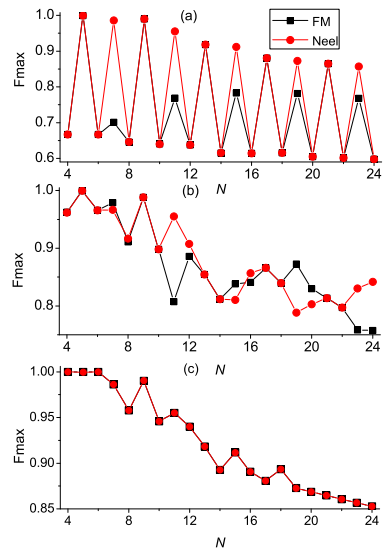}
\caption{(Color online) Comparison of the maximum average fidelity that can
be achieved as a function of chain length \emph{N} for different initial
state configurations; the FM ground state and the N\'{e}el state. The
results are obtained within the time interval [0,500/$\left\vert
J\right\vert $ ] and for field strengths (a) \emph{h}=0.0, (b) \emph{h}=1.0,
(c) optimal field strengths.}
\label{fig:3}
\end{figure}

In Fig.3 we compare the communication fidelity when an arbitrary qubit
state is transferred through two initial channel configurations of various
length. One of these channels is chosen to be the ground state $\left\vert
\mathbf{0}\right\rangle$ of a FM chain while the other channel corresponds
to the N\'{e}el state. In a single excitation subspace, for an
unmodulated XY Hamiltonian with length $N$, it has pretty good state
transfer (PGST) between the two ends of the chain if and only if $N=p-1$ or
$N=2p-1$, where $p$ is prime, or if $N=2^{m}-1$ \cite{Godsil}. Here
PGST means that for every $\epsilon>0$, there exists time $t$ such that
the maximum fidelity is greater than $1-\epsilon$. For example, when
$N=6$, Tmax$\approx298$, and Fmax$>0.99$ ($\epsilon=0.01$) \cite{Godsil}.
But for a long chain, Tmax will be very long for $\epsilon=0.01$. As
most works about quantum state transfer, we choose a finite time
window [0, 500/$\left\vert J\right\vert $], in which we can obtain a MAF for $N=7, \epsilon=0.02$. Taking a
finite Tmax is physically reasonable, because the receiver can not
wait for a very long time. We set fixed magnetic field \emph{h}=0.0 (\emph{h}=1.0) for plots
Fig. 3(a) (Fig. 3(b)).  First notice that when \emph{N}=5, both channels
yield Fmax $\approx 1$ indicating that near perfect state transfer can be
realized. Secondly, in the absence of a magnetic field (Fig. 2(a)) the MAF
associated with the N\'{e}el channel is typically greater than or equal to
the values which occur for the FM ground state configuration. The figure
suggests that for \emph{h}=0.0 the N\'{e}el channel supports a better state
transfer than $\left\vert \mathbf{0}\right\rangle$ for chains containing $%
7+4n$ $(n=0,1,2...)$ sites. In Fig. 3(b) we compare the results when an
external magnetic field is present. In this case the MAF associated with the
N\'{e}el state can be lower than the $\left\vert \mathbf{0}\right\rangle$
channel for certain $N$. We also find that when $N=5+4n$ and $N=6+4n$ $%
(n=0,1,2...)$, the MAF is equal for the N\'{e}el channel and the FM ground
state channel. In Fig.3(c) we plot the MAF for optimal choice of magnetic field.
Through numerical calculation we find that the MAF and the time Tmax
at which the average fidelity gains its maximum value (Fig.4(c)) are always equal
when using the the FM ground state and the N\'{e}el state as the initial state.
Then half excitations of the chain length and a single excitation
shows same transmission quality when considering optimal choice of
magnetic field. Fig.3(c) also shows that the MAF are greatly enhanced compared with
the fixed magnetic field which is plotted in Fig.3(a),(b). $N=4,5,6$
gives nearly perfect (Fmax=0.999) state transfer.

In Fig.4 we plot the time Tmax as a function of site number $N$. In the absence of a magnetic field (Fig.4(a)) we find that
when Fmax is equal for chains differing in length by one unit the associated
arrival times Tmax are also equal. In the presence of a magnetic field (Fig.
4(b)) this behavior also exists. For certain chain lengths a shorter arrival
time accompanies the higher MAF which can be obtained using the N\'{e}el
channel, e.g., \textit{N} =11, 19 (\textit{h}=0.0) and \textit{N}=12, 16 (%
\textit{h}=1.0).

\begin{figure}[tbph]
\centering
\includegraphics[scale=1.0,angle=0]{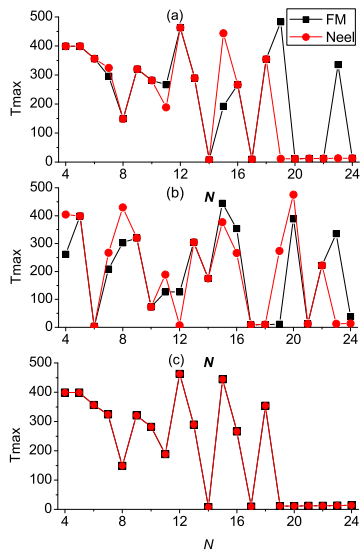}
\caption{(Color online) The time Tmax at which the average fidelity gains
its maximum. (a) h=0.0, (b) h=1.0 (c) optimal field strengths $h\in[0,2.0]$.}
\label{fig:4}
\end{figure}

Since the N\'{e}el state is not an eigenstate of the AFM chain there is a
large probability that it will collapse to another state when we attempt to
obtain it through measurement. These other possible states will contain the
same number of excitations as the N\'{e}el state but the location of the
excitations will generally be different. For instance, when \textit{N}=6 and
\textit{M}=3 the locations of the excitations for the N\'{e}el state are at
sites 2, 4, and 6. Suppose these excitations occupied other sites, say sites
2, 3, and 4. We now check to see how the average fidelity is affected by
such a re-ordering. We sample random configurations for chains containing
\textit{N}=6 and \textit{N}=15 sites having \textit{M}=3 and \textit{M}=7
excitations respectively. For the \textit{N}=6 site chains we select the
excitation locations to be (2, 3, 4), (3, 4, 5), (2, 4, 6), and for the
\textit{N}=15 site chains we choose (3, 4, 6, 10, 11, 12, 14), (2, 3, 7, 8,
10, 11, 13). Using the arrival times Tmax associated with the N\'{e}el
ordered states we calculate the difference in the average fidelity between
these different configurations. For \textit{h} = 1.0 we find a difference of
only $5.55\times10^{-16}$ between the orders (2, 3, 4) and (3, 4, 5). A
comparison between orders (2, 4, 6) and (2, 3, 4) yields an even smaller
difference for the same value of \textit{h}. For \textit{h}=0.0 and \textit{N%
}=15 the difference in the fidelity for the two configurations above is $%
3.55\times10^{-15}$. We have also checked other initial state configurations
using various values for \textit{N} and find that the average fidelity is
nearly equal to the corresponding \textit{N}-site N\'{e}el state at the same
time which maximizes the N\'{e}el channel average fidelity. We conjecture
that when the number of excitations is roughly similar to half of the system
size, the evolution of the average fidelity only depends on the number of
excitations in the chain and is independent of their ordering. If this
prediction holds, the initialization process of the AFM chain would be
simplified. If the chains state collapses via measurement to any state
containing a fixed and known number of excitations we could predict the
behavior of the subsequent evolution of the fidelity.

\section{CONCLUSIONS}

In this work we have shown that multi-excitation channels can provide
suitable pathways for quantum communication. Some of the AFM chains we have
considered have been found to outperform state transfer protocols based on
ferromagnetic media which are initialized to the ground state. Specifically,
we have found certain N\'{e}el state configurations which allow a quantum
state to be transmitted in a shorter amount of time and arrive with a higher
average fidelity than in the FM case. Moreover, numerical calculations
support our conjecture that the quality of state transfer through a
multi-excitation subspace only depends on the number of excitations present
in the initial state of the system. Since the fidelity of state transfer
appears to be independent of the ordering of the initial excitations, we
believe that the AFM ground state can serve as a communication channel.

These results should be interesting to test experimentally, perhaps using
NMR methods \cite{Takigawa1996}, fabricated AFM nano-chains \cite%
{Hirjibehedin2006}, or optical lattices \cite{Simon2011,Hofstetter2006}.

\section*{ACKNOWLEDGMENTS}

This material is based upon work supported by the NSF (Grant No's. 11005099,
60677044) and by Fundamental Research Funds for the Central Universities
(Grant No. 201013037).

\end{document}